# Dynamic cloaking effect of a vector barrier in bilayer graphene

S. Maiti[1], R. Biswas[2,*] and C. Sinha[2,3]

[1]Ajodhya Hills G. S. A. High School, Ajodhya, Purulia, West Bengal-723152, India

[2]P. K. College, Contai, Purba Medinipur, West Bengal- 721401, India

[3]Indian Association for the Cultivation of Science, Jadavpur, Kolkata-700032, India

**ABSTRACT:**

Transmission profiles in bilayer graphene have been studied theoretically in presence of a pair of delta function magnetic barriers. Two types of asymmetric Fano resonances are discussed in connection to the electronic cloaking effect in graphene nanostructures. One arises in a static (time independent) vector barrier due to the quantum interference between the discrete hole like state inside the barrier and the electron like continuum outside. In this case the computed results for normal incidence indirectly verify that the cloaking effect is a manifestation of the chirality conservation in charge transport through bilayer graphene scalar barriers. While the other arising due to discrete continuum coupling via the exchange of photon under the application of an external time periodic scalar potential. The study of Fano resonances in transmission spectrum is very much urgent in regards of localization of charge carriers in graphene nano structures for its application in digital device fabrication.

*E-Mail: rbiswas.pkc@gmail.com

# 1. Introduction:

Since the experimental realization of graphene [1], a wide variety of research work appeared in the literature regarding both the static and dynamic properties of mono and bilayer graphene, particularly due to many of their exhotic properties [2]. Unlike the monolayer graphene (MLG), chiral charge carriers in the bilayer graphene (BLG) are massive and obey quadratic disperssion [2-6], similar to those of the Schroidinger electrons. The chiral property arises due to the valley degrees of freedom defined by the pseudo spin of the Dirac fermions. The eigen values of the chirality operator that corresponds to the alignments between the pseudo spin and the momentum is responsible for the appearence of the electron and hole like states that coexist in the vicinity of the Dirac point [6]. The conservation of chirality leads to the phenomenon of Klein transmission resulting in perfect transmission in MLG [7] but almost no transmission for BLG at normal incidence [8] of low energy charge carriers on an electrostatic barrier. Similar to the case of transmission of the Schroidinger electrons, the probablity of transmission of the Dirac fermion through a BLG electrostatic barrier increases exponentially with the increase in incident energy of the charge carriers [9]. While, the most important difference is that, the transmission spectra of the massive Dirac fermion for near normal incidence exhibit the asymmetric Fano resonances (FR) arising due to the coupling between the discrete hole like state within the barrier and the electron like continuum outside [9, 10]. A dramatic change occurs at normal incidence where the FRs are not at all observed because of the decoupling between the discrete and the continuum, the discrete state being cloaked by the electrostatic barrier in order to maintain the conservation of chirality in the BLG. After the reporting [9-11] of this electronic cloaking effect (analogus to the optical cloaking effect), several authors attempted theoretically [12] to use this effect in the context of carrier confinement in the BLG. Recently the experimental evidence of this electronic cloaking effect is reported due to Lee et. al. [13] by probing the phase coherent transport behavior of a dual gated BLG transistor. Further, it is shown that [14] the cloaked states could decay slowly into the continuum via the trigonal warping effect. Since the cloaking effect is a manifestation of the chirality conservation and consequently the time revarsal symmetry in the system, it is quite natural to disappear this effect in case of magnetic barriers, due to the broken chirality in the vector potential barrier in a BLG. Although a few reports exist in the literature on the transmission property of the massive Dirac fermion in BLG magnetic barriers [15-18], the appearence of the FR and the destruction of the cloaking effect are not discussed so far. The study of the FR in BLG is highly needed to tailor the electron-hole coupling during transmission of the charge carriers either by controlling the angle of incidence or the strength of the external magnetic field.

The BLG is a good candidate for nano-electronics, where the characteristics of the charge carriers could be controlled by an applied time dependent bias between the two layers. Study of the transport phenomena of externally driven Dirac fermions is still in high demand for the control of charge and spin transports in graphene based nano devices. Further, it is reported very recently [19] that the tunnelling between graphene Landau levels driven by the time periodic perturbation is theoretically linked to the process of pair creation from vacuum, a prediction of quantum electrodynamics (QED) [20]. Particularly, the study of the quantum transport in periodically driven system is interesting in the sense that the quantum interference within an oscillating time periodic potential gives rise to additional Floquet sidebands arising from the absorption or emission of photons from the driving field. Applying Tien-Gordon approach [21], it is shown that the conductance interference pattern could be modified by tuning AC gate voltage [22]. In a further work [23, 24], it is shown that the Fabry Perot resonance and sharp Fano anti-resonances in the electronic conductances in BLG flake created under magnetic field are suppressed partially or fully by the application of the ac voltage. Finally a question arises whether the cloaking effect would servive or not by any means in presence of the oscillating potential. Recently, we reported theoretically [25] different quantum interference effects e.g., beating oscillations in Fabry – Perot resonances and two types of asymmetric Fano resonances occurring in the time periodic modulation of the magneto-transmission in monolayer graphene. This motivates us to extend our model to a bilayer graphene (BLG) that could be a potential candidate for the practical applications in the field of magnetic nano-fabrications. Although very recently the application of time periodic potential on BLG electro static barrier is studied theoretically [26], the modification, if any, of cloaking effect was not discussed.

The present study addresses the electronic transport through a pair of oppositely directed delta function magnetic barriers in a bilayer graphene in presence (and absence) of an external time periodic scalar potential. The time dependent problem is solved in the framework of a non-perturbative Floquet formalism [27, 28]. Two types of FR's are discussed, one arising due to the direct coupling between a discrete hole like state and a electron like continuum, while the other arising due to indirect coupling via the emission or absorption of photons. Here we have shown that the disappearance of the FR (a manifestation of the cloaking effect) might occur by the application of a time periodic potential on a BLG vector barrier. Unlike the electrostatic barrier in BLG (where cloaking effect is noted for normal incidence only), here in presence of a time periodic potential, the cloaking effect has been observed (due to either emission or absorption of photons) for the normal as well as glancing incidences.

## 2. Theoretical Formulation:

We consider a sheet of BLG lying in the xy plane (where the atoms are arranged in Bernal stacking [3]) subjected to a static vector potential barrier of width '$L$' producing a pair of oppositely directed delta function magnetic field (of equal in strength and separated by a distance '$L$') acting normal to the xy- plane and uniform along the y- direction as shown in fig. 1. The low energy excitations can be given by the Hamiltonian

$$H_0 = g \begin{pmatrix} 0 & \pi_-^2 \\ \pi_+^2 & 0 \end{pmatrix} \qquad (1)$$

where $\pi_+ = p_x + i(p_y + eA_y)$ and $\pi_- = p_x - i(p_y + eA_y)$. Here, $p_{x,y} = -i\hbar \partial_{x,y}$ and $g = 1/2m$, where the effective mass $m = 0.043 m_e$, $m_e$ being the rest mass of electron.

The vector potential $\vec{A} = (0, A_y(x), 0)$ that corresponds to a pair of delta function magnetic barriers satisfies the following relation [29]

$A_y(x) = \beta$ (in units of $B_0 l_0$) for $0 < x < L$ (region-II)    2(a)

   $= 0$ elsewhere (for region I, $x < 0$ and for region III, $x > L$)

and $\vec{B}(x) = \pm \beta \, \hat{z}$ (in units of $B_0$) for $x = 0$ (upper sign) and $x = L$ (lower sign)

    $= 0$   elsewhere .             2(b)

Here $l_0 = \sqrt{\dfrac{\hbar}{eB_0}}$ is the length scale with a typical magnetic field strength $B_0$; The above potential profile can be created by depositing a ferromagnetic strip on top of the BLG.

   Now the solution of the Dirac equation corresponding to the Hamiltonian eqn. (1) can be given by two component pseudo-spin wave functions

$\varphi_a(x, y) = \left[ A e^{ik_b x} + B e^{-ik_b x} + C e^{\lambda_b x} + D e^{-\lambda_b x} \right] e^{ik_y y}$    3(a)

$$\varphi_b(x,y) = \left[Ae^{ik_bx+2i\theta_b} + Be^{-ik_bx-2i\theta_b} + \mu Ce^{\lambda_b x} + \eta De^{-\lambda_b x}\right]e^{ik_y y} \qquad 3(b)$$

where $k_b^2 = |E| - (k_y + \beta)^2$, $\lambda_b^2 = |E| + (k_y + \beta)^2$, $\theta_b = Tan^{-1}\left[(k_y + \beta)/k_b\right]$,

$\mu = -(\lambda_b - k_y - \beta)^2/E$ and $\eta = -(\lambda_b + k_y + \beta)^2/E$. $A, B, C$ and $D$ are the constant coefficients. All the parameters are written in dimensionless unit, e.g., $x \to x\, l_0$, $\beta \to \beta B_0$, $E \to EE_0$, $V_0 \to V V^{/}$ and $\omega \to \omega \omega_0$, where $\omega_0 = \frac{v_F}{l_0}$, $E_0 = \hbar\omega_0$ and $V^{/} = \frac{E_0}{e}$.

Whereas, in presence of the oscillating potential the time dependent Hamiltonian takes the form $H = H_0 + V_0 Cos\omega t$ and the corresponding solution of the time dependent equation of motion is given by

$$\Psi(x,y,t) = \sum_{l,m=-\infty}^{\infty} \begin{pmatrix} \varphi_a^m(x) \\ \varphi_b^m(x) \end{pmatrix} e^{ik_y y} J_{l-m}(\alpha) e^{-i(Et+l\omega)} \qquad (4)$$

Here $J_{l-m}(\alpha)$ is the Bessel function of order '$l$-$m$', '$l$' being the index of photon number (absorbed /emitted), '$m$' being the sideband index and $\alpha = V/\omega$.

Finally, matching the wave functions and their derivatives at the two interfaces $x = 0$ and $x = L$ one can calculate the transmission coefficients $T_m$ at different Floquet sidebands corresponding to energies $E_m = E + m\omega$.

$$T_m = \frac{cos\theta_m}{cos\theta_0} \left|\frac{D_m}{F_0}\right|^2 \qquad (5)$$

with $\theta_m = \tan^{-1}\left(\frac{k_y}{k_x^m}\right)$ and $(k_x^m)^2 = (E_m)^2 - (k_y)^2$ ; $F_0$ is the amplitude of the wave incident at an angle $\theta_0$ in the region I.

3. **Results and discussions:**

In order to show that the tunneling spectra for a static BLG vector barrier also exhibits the characteristic FR's (similar to an electrostatic barrier), we display in Fig. 2, the total transmission ($T_c$) for four different values of the y-component of momentum as $k_y$= 0, -0.01, -0.02 and -0.05. Here 'B', 'L', 'E' (or '$V_0$'), and 'ω' are expressed in units of $B_0$ (say, 0.1 T), $L_0$ (= 81.13 nm), $E_0$ (=8.113 meV), and $\omega_0$ (=12.325×10$^{12}$ Hz) respectively in all the figures. The salient point to be noted is the following. In a conventional heterostructure, the transmission spectrum exhibits the resonant tunneling behavior which is Lorentzian in shape and arises due to the coherent superposition of states within the well, while the present transmission spectrum exhibits the characteristic Fano resonances to be described below. From the figure it is clear that the asymmatric Fano Resonances, i.e., a sharp minimum followed by a maximum (unlike the Lorentzian), appear for the case of normal ($k_y$=0) as well as glancing incidences on a vector potential barrier. For the present parameters, the characteristic FR arises due to the presence of a discrete hole like quasi bound state around the energy $E_b$ ~ 0.12 inside the vector barrier. The existence of the FR at normal incidence implies that even for such case, the pseudo spin flip is possible at the interface of two regeions having different vector potential, since the chirality is not a good quantum number for a system with broken time reversal symmetry. Thus the cloaking effect (non-appearance of the FR) exhibited by a scalar barrier (unlike the static vector barrier) is a consequence of the conservation of chirality in BLG. The shape of the FR is quite sensitive to the magnitude of the y-component of the momentum of the incident charge carrier. Particularly, for $k_y > -0.02$, a sharp minimum is followed by a sharp maximum while reverse is the case for $k_y < -0.02$. The reason may probably be due to the change in the position and width of the quasi-bound hole like state inside the barrier.

Next, to study the effect of the external time harmonic potential on the quantum interference effect (the Fano transmission) in BLG vector barrier, we display in Fig. 3(a) the total transmission ($T = \sum_m T_m$) for three different values of the amplitude of the oscillating potential, e.g. $V$= 0.0, 0.5 and 1.0 for a particular frequency $\omega$ = 0.01. It is interesting to note that with the increase in amplitude $V$, the peak to dip ratio (PDR) of the FR (at $E_b$= 0.12) corresponding to the static barrier decreases and finally the FR disappears beyond a particular value of $V$ (> 1.0). This disappearence of the FR (cloaking efect) is somewhat similar to the well established cloaking effect of the electrostatic barrier [9-11] at the normal incidence in BLG, however, the cause may be different. The origin for the present cloaking effect could be attributed to the fact that, when the incident energy equals that of the discrete quasi bound hole like state inside the barrier, the photon exchange channels are more probable than the no photon proscess and the charge carriers propagate evanescently through the vector barrier. However, for incident energies greater or less than $E_b$, the incident electron may transmit with higher probablity due to photon exchange processes than the nophoton one via resonant transmission through the barrier. This photon assised cloaking effect is found to occur both for the normal as well as for the glancing incidence on the vector barrier.

The total transmission, for energies away from the resonant ones ($E = E_b$), remains almost unaffected at lower amplitude of the oscillating potential. The effect may be appreciable at higher value of the incident energy. Fig.3(b) also reveals that, for the single photon process, the width of the FR increases and the position of the FR is slowly blue shifted with the increase in $V$, whereas for the two photon case, the peak to dip ratio increases appreciably.

The main characteristic feature of the photon assisted resonant tunneling is the appearance of the FR's, the positions of which depend particularly on the frequency of the oscillating potential and the energy of the quasi bound state, governed by the energy conservation relation $E = E_b + n\omega$. As is noted from Fig. 3(c), for two different values of the frequency, (e.g., $\omega_1$ & $\omega_2$), the shift in the position of the FR is equal to the difference ($\omega_1$-$\omega_2$). The figure also exhibits that for each ω, another short FR with smaller PDR is noted at higher value of the incident energy. It means that for transmission associated with higher number of photon exchange is less probable than that for the lower one, as was also noted earlier [28]. Thus the BLG vector barrier could be used as a frequency detector in the THz range.

To study the effect of the y-component of momentum on the photon induced FR's, we display in Fig.4(a) the total transmission for three different values $k_y$, e.g., -0.9, -1.1 and -1.3 for V= 0.6 and ω = 0.6. Fig. 4(a) reveals that the threshold energy of finite transmission increases with the increase in the magnitude of $k_y$, in view of the energy conservation relation $k_x^2 = |E| - k_y^2$. Further, the position of the photon induced FR moves towards the higher energy end with the increase in $|k_y|$, particularly due to the shift in the position of the quasi bound state inside the vector barrier. On the contrary, the width of the FR decreases and the secondary PDR (on either side of the main peak and dip) increases with the increase in $|k_y|$. Whereas, the primary PDR becomes independent of $|k_y|$ for both the single and two photon processes.

Finally, Fig.4(b) depicts the total transmission for three different barrier widths, $L$ = 1.2, 1.4 and 1.6 with V= 0.6 and ω = 0.6. With the increase in barrier width, the photon induced FR moves towards the lower energy end following a systematic decrease in the resonance width and finally, the photon induced FR disappears at a larger width, probably due to the non-survival of the quasi bound state within the barrier.

4. **Conclusion**:

Time periodic perturbation is used to control the chiral tunneling property of charge carriers through a pair of delta function magnetic barriers in bilayer graphene. Through the calculation of the transmission coefficient we have shown that such a vector barrier under static condition does not show any electronic cloaking effect. With the use of non-perturbative Floquet method we have also computed the transmission in different Floquet sidebands arising due to a sinusoidal scalar time dependent potential. It is shown that the Fano resonance in static BLG vector barrier may disappear (also termed as cloaking effect in case of a static BLG scalar barrier at normal incidence) due to exchange of photon under the application of a sinusoidal scalar potential. This photon induced cloaking effect reported here is for the first time in the literature.

Figure Captions:

Fig.1: (a) Vector potential profile (A(x) v.s. x) corresponding to a pair of oppositely directed δ-function magnetic barriers of strength 'B' and separated by a distance 'L'. (b) Magnetic field profile corresponding to the vector potential as shown in Fig.(a). (c) Sinusoidally varying time dependent scalar potential of amplitude 'V' and frequency 'ω' applied in region II (the dash line represents the value at any instant of time 't').

Fig.2: (Color online only) Transmission coefficient '$T_c$' plotted as a function of incident energy 'E' for a time independent vector barrier with 'B' = 1 and 'L' = 1. Solid (black) line for '$k_y$'= 0, dash (red) line for '$k_y$'= -0.01, dot (blue) line for '$k_y$'= - 0.02 and dash-dot (green) line for '$k_y$'= -0.05.

Fig.3: (Color online only) Total side band transmission $T_c$ ($\sum_m T_m$) plotted as a function of incident energy 'E' for a time dependent vector barrier with 'B' = 1 and 'L' = 1. (a) for '$k_y$'= -0.02 and 'ω'=0.01. Solid (black) line for 'V'= 0, dash (red) line for 'V'= 0.05 and dot (blue) line for 'V'= 0.1; (b) for '$k_y$'= - 1 and 'ω'=0.5. Dot (black) line for 'V'= 0, solid (blue) line for 'V'= 0.5 and dash (red) line for 'V'= 1; (c) for '$k_y$'= - 1 and 'V'=0.7. Solid (black) line for 'ω'= 0.7, dash (red) line for 'ω'= 0.6 and dot (blue) line for 'ω'= 0.5.

Fig.4: (Color online only) Same as Fig.3 but (a) for 'L' = 1, 'V'= 0.6 and 'ω'= 0.6. Solid (black) line for '$k_y$'= -0.9, dash (red) line for '$k_y$'= -1.1 and dot (blue) line for '$k_y$'= -1.3; (b) for '$k_y$'=-1, 'V'= 0.5 and 'ω'= 0.7. Solid (black) line for 'L'= 1.2, dash (red) line for 'L'= 1.4 and dot (blue) line for 'L'= 1.6.

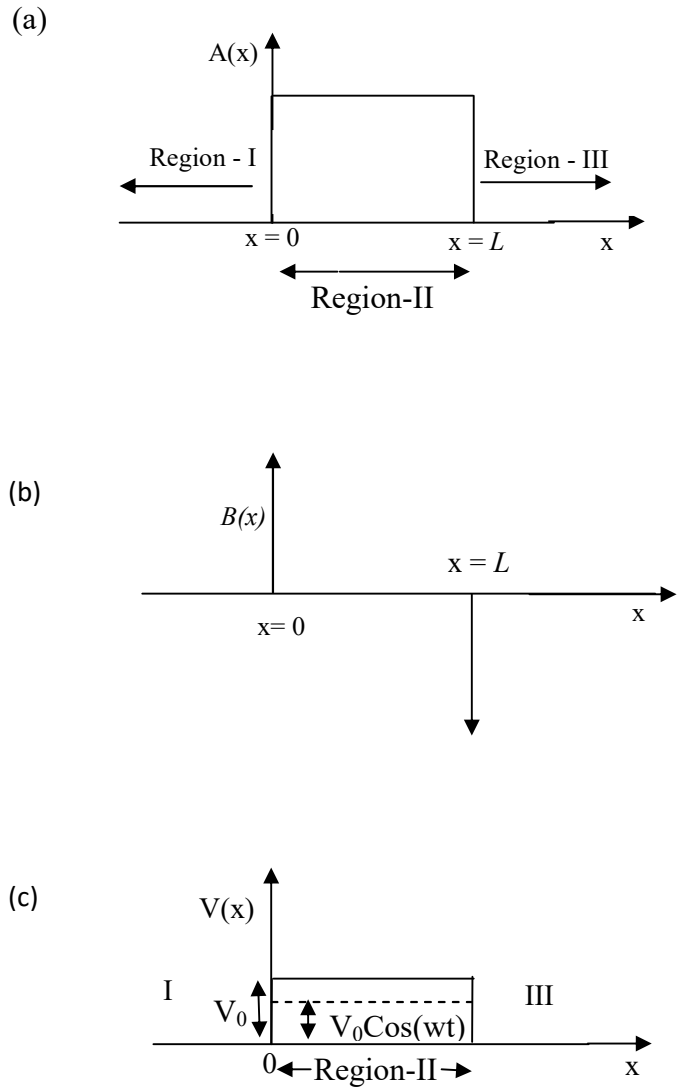

Fig.1

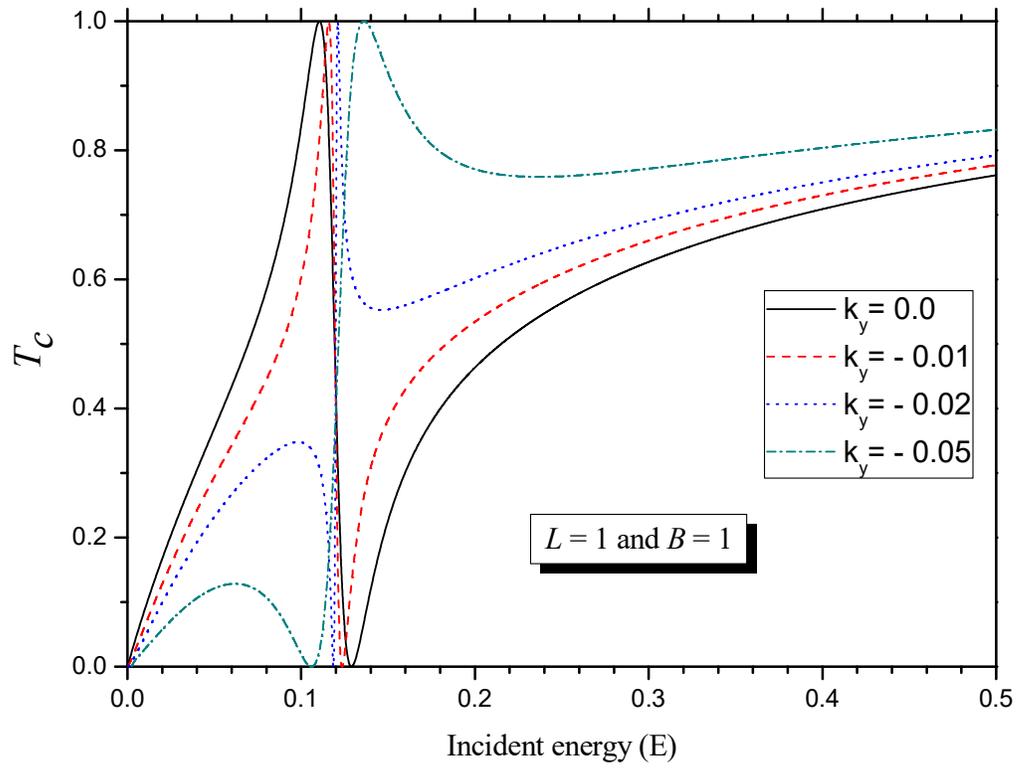

Fig.2

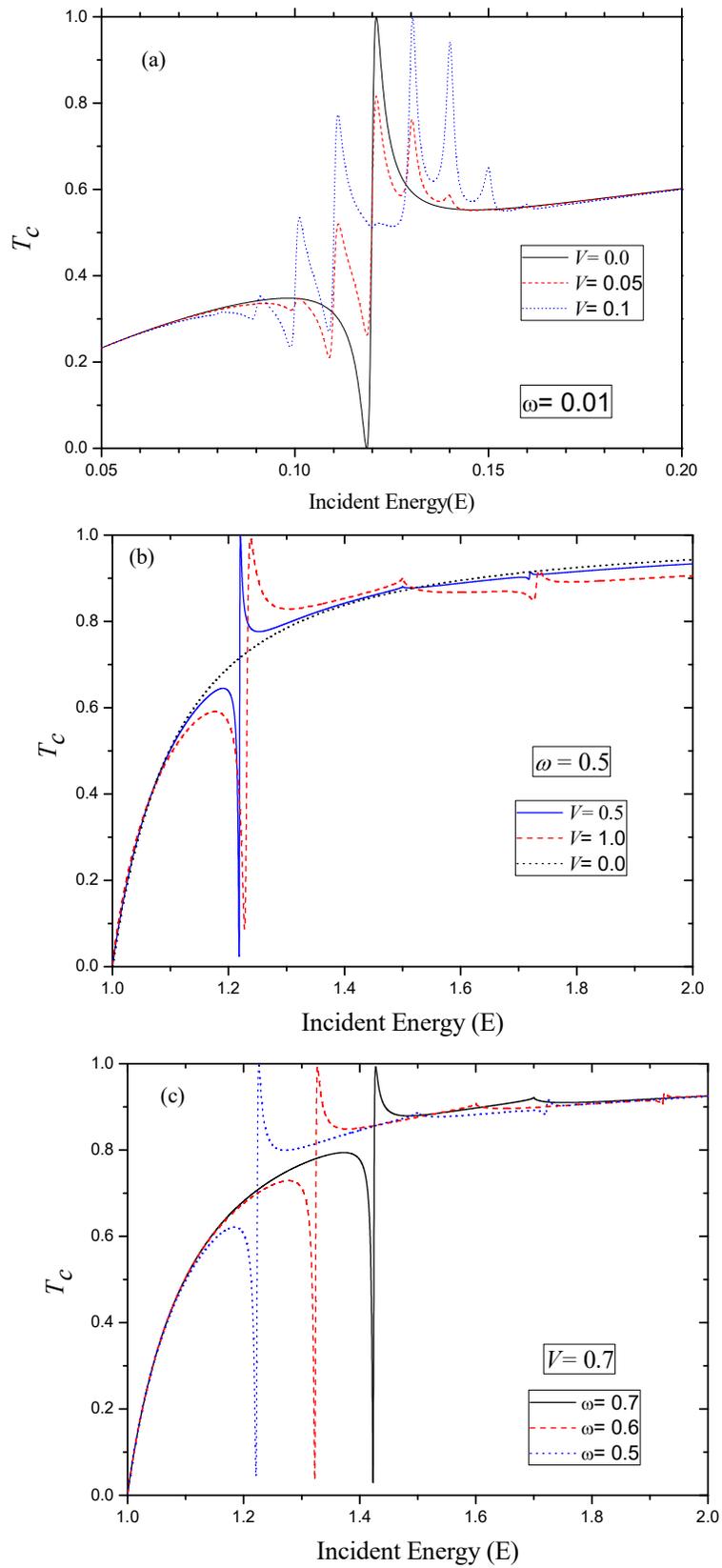

Fig.3

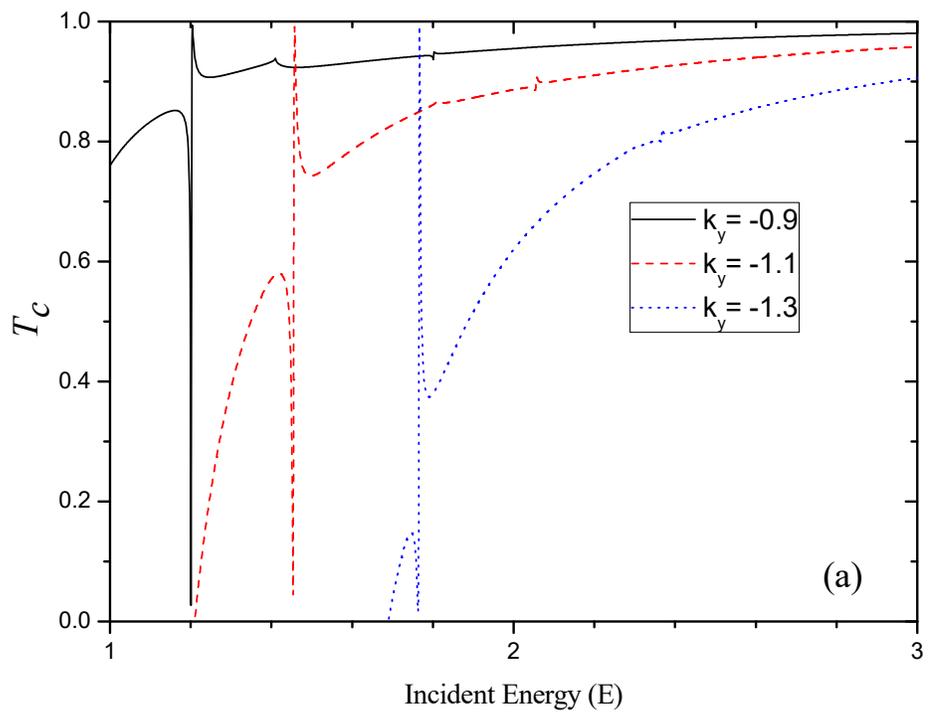

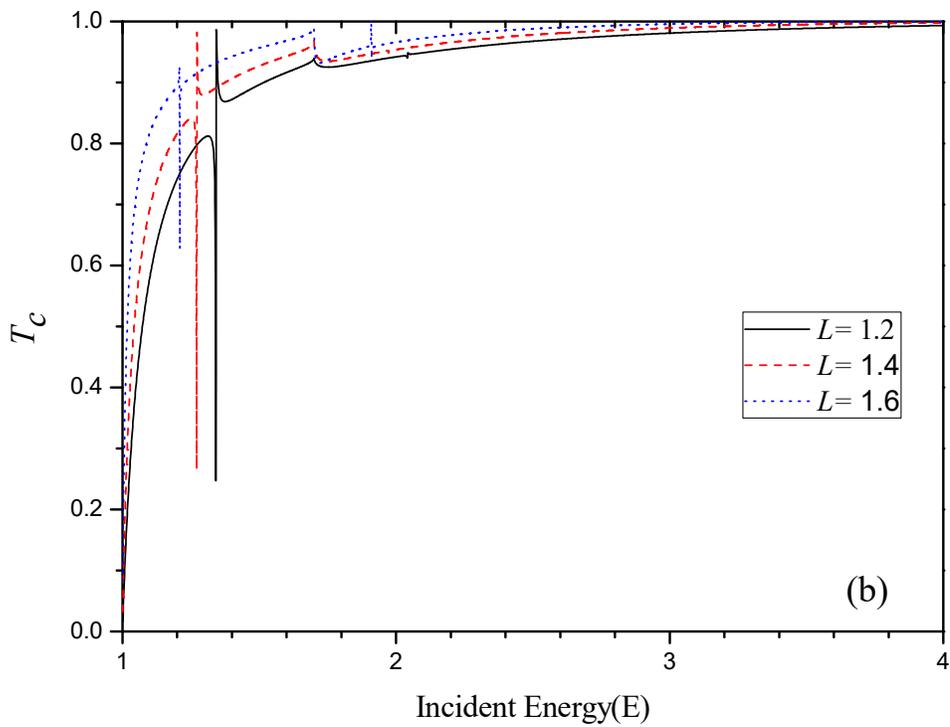

Fig.4